# Hadron Colliders and Hadron Collider Physics Symposium

Dmitri Denisov

Fermi National Accelerator Laboratory, Batavia, Illinois 60510 USA

**Abstract.** This article summarizes main developments of the hadron colliders and physics results obtained since their inception around forty years ago. The increase in the collision energy of over two orders of magnitude and even larger increases in luminosity provided experiments with unique data samples. Developments of full acceptance detectors, particle identification and analysis methods provided fundamental discoveries and ultra-precise measurements which culminated in the completion and in depth verification of the Standard Model. Hadron Collider Physics symposium provided opportunities for those working at hadron colliders to share results of their research since 1979 and helped greatly to develop the field of particle physics.

## 1 Hadron Colliders

The idea of using colliding beams to increase center of mass energy was developed in 1960's with $e^+e^-$ colliders designed and constructed. This idea is based on the kinematics of the collision. In the case of head on collisions for colliding beams of the same mass particles the center of mass system and the laboratory system coincide and the total center of mass energy is twice the energy of the beams. In the case of fixed target experiments, which provided critical experimental information at the early stages of the Standard Model developments, for proton proton collisions the center of mass energy is $\sim \sqrt{(2m_p E_{beam})}$ in the approximation of beam energy $E_{beam}$ well above mass of proton $m_p$. This square root dependence reduces the center of mass energy of the fixed target experiments substantially. For example, at 1 TeV beam energy fixed target center of mass energy is ~50 GeV, while two 1 TeV beams colliding provide 2 TeV in the center of mass or 20 times larger center of mass energy. Only particle colliders are able to create new elementary particles with masses up to the top quark mass of 173 GeV [1].

Fig. 1 illustrates history of the particle colliders since early 1960's. Among different types of particle colliders, proton-proton and proton-antiproton colliders, provide highest center of mass collision energies which led to the most important discoveries in the particle physics over last thirty years.

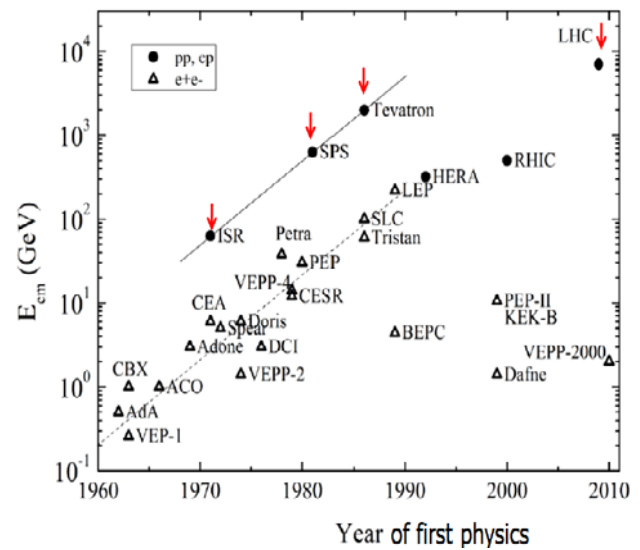

**Fig. 1.** Center of mass energy of particle colliders between 1960 and 2010. Year is the first year of physics operation of a collider. Figure is adopted from Ref. [2].

## 2 Hadron Colliders Physics

Main areas of the hadron collider physics studies were rapidly changing between first proton-proton collider Intersecting Storage Rings (ISR) went into operation in 1971 and Large Hadron Collider (LHC) started its operation in 2009. These studies reflected most interesting and urgent developments in particle physics and provided experimenters with data to perform critical experiments leading to exciting discoveries.

Majority of the studies at hadron colliders could be separated into two main areas: search for new particles and interactions predicted in the Standard Model or theories beyond the Standard Model and precision measurements of the elementary particles properties and their interactions. Among fundamental questions under study at the hadron colliders are existence of the structure inside quarks, origin of the mass, reason for the observed matter-antimatter asymmetry, origin of the cosmic dark matter and understanding of space-time structure at the smallest possible for experimental studies distances and time intervals.

Among most fundamental discoveries made at the hadron colliders, discoveries of W and Z bosons, top quark and recently Higgs boson [3] are all excellent examples. Ultra-precise measurements, such as of W boson mass [4] with accuracy of 15 MeV for a particle with 80 GeV mass, are absolutely critical to verify predicting power of the Standard Model and to measure its fundamental parameters.

## 2.1 Intersecting Storage Rings

ISR collider was the first hadron collider which operated at CERN from 1971 to 1984. This collider center of mass energy of up to 56 GeV was equivalent to a fixed target experiments with ~2 TeV beam energy – not available for experiments even today. This pioneer hadron collider was critical to develop and understand principles of proton beams storage and collisions, achieving very high vacuum, and solving multiple accelerator challenges. Main parameters of the ISR collider are listed in Table 1.

**Table 1.** Main parameters of the ISR collider at CERN.

TABLE 1. *Main parameters of the ISR*

| | |
|---|---|
| Number of rings | 2 |
| Circumference of rings | 942.66 m |
| Number of intersections | 8 |
| Length of long straight section | 16.8 m |
| Intersection angle at crossing points | 14.7885° |
| Maximum energy of each beam | 28 GeV |
| Hoped for luminosity (per intersection) | $4 \times 10^{30}$ cm$^{-2}$ s$^{-1}$ |
| *Magnet (one ring)* | |
| Maximum field at equilibrium orbit | 12 kG |
| Maximum current to magnet coils | 3750 A |
| Maximum power dissipation | 7.04 MW |
| Number of magnet periods | 48 |
| Number of superperiods | 4 |
| Total weight of steel | 5000 tons |
| Total weight of copper | 560 tons |

Experiments at the ISR accelerator covered many hot topics of the 1970's particle physics. Total cross sections of proton-proton interactions were measured up to very high energies and un-expected increase in these cross sections with increase in beam energy was confirmed. Elastic cross sections have been measured with high precision as well as yields of many elementary particles. Studies of quark-quark, quark-gluon and gluon-gluon scattering have started during final generation of ISR experiments. Detection of jets created by quarks and gluons and their studies helped to quickly develop just introduced quantum chromo-dynamic. Very important were the developments of the experimental methods to perform experiments "inside the accelerators" as colliders interaction regions are heavily populated by machine components and suffer substantial backgrounds from beam losses.

1970's was the decade of major particle physics discoveries such as c- and b-quarks, tau-lepton and others. Why these fundamental discoveries have been made at fixed target proton accelerators and e$^+$e$^-$ colliders, while ISR energy and luminosity provided an opportunity to produce these particles? Most natural answer on this question is that all these experimental discoveries were to large extent un-expected (no specific theoretical predictions) and experimental methods for fixed target beams were better developed with larger and more sophisticated experiments. At the same time e$^+$e$^-$ colliders, due to point-like beams, did provide extremely precise scans vs center of mass energy to detect production of J/ψ particle which consists of c-quark and c-anti quark. But the developments of the ISR accelerator and detectors were so successful that over next three decades all new elementary particles have been discovered at hadron colliders.

## 2.2 Super Proton Synchrotron

By late 1970's Standard Model developments were progressing extremely quickly and prediction of new bosons, W and Z, carriers of weak interaction has been made. Based on the theoretical predictions and available experimental data mass of these bosons was expected to be ~100 GeV. Fixed target beam with energy of ~5 TeV was required to achieve such high masses, far beyond energies of accelerators available at that time.

An excellent idea helped to address pressing physics topic of W and Z bosons existence quickly: by converting existing circular accelerators of a few 100's GeV energy into colliders with protons and anti-protons circulating in the same ring and beam pipe in the opposite directions. Exactly the same mass of the antiproton and opposite electric charge makes this idea feasible. Most serious challenge was to make large number of antiprotons and even more challenging "cool them" into a small diameter beam to collide with high interaction rate with a beam of protons circulating in the opposite direction.

Development of the first antiproton source at CERN was triumph of accelerator science. Fig. 2 presents a photo of the antiproton source during its construction at CERN. At the same time detectors for the Super Proton Synchrotron (SPS) collision regions have been developed. One of them, UA1 detector, resembles modern collider detectors closely: it has central tracking detector to measure tracks of the charged particles, hermetic calorimetry as well as muon system to detect muons which penetrate large amount of material. The detector covered close to $4\pi$ solid angle around interaction region and provided and excellent environment to measure decays of heavy particles created in the center of the detector.

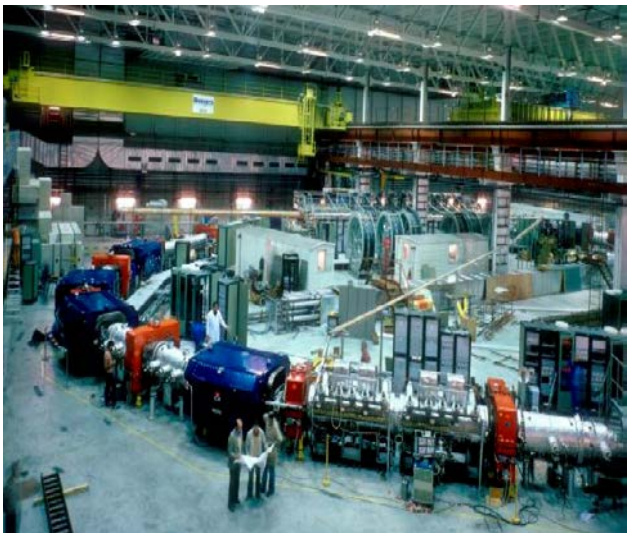

**Fig. 2.** Construction of the first antiproton source at CERN.

SPS collider started operation in 1981 and reached 630 GeV center of mass energy which provided experimentalists with an opportunity to detect production and decay of W and Z bosons [5]. Fig. 3 presents discovery of the Z boson in the decay mode to 4 muons with a mass of ~90 GeV.

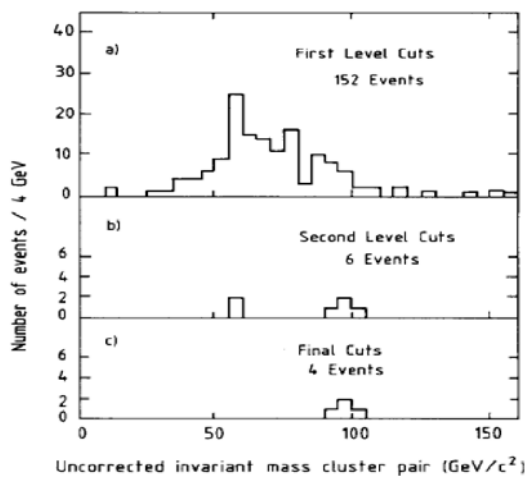

**Fig. 3.** Invariant mass spectrum of two muons by UA1 collaboration.

## 2.2 The Tevatron

On the way to even higher collision energies the limiting factor at SPS was 2 T magnetic field iron saturation in the bending magnets in the accelerator ring. In order to increase energy of the collider even further either larger diameter ring or higher field magnets were required. At Fermilab the proposal to increase bending field by using superconducting magnets with 4.5 T field has been developed and implemented in early 1980's in the design of the Tevatron, first accelerator to reach 1 TeV per beam energy and collision energy for proton-antiproton collider mode operation of 2 TeV.

Tevatron started operation in 1985 with first high luminosity run starting in 1992 with two modern, general purpose detectors CDF and D0 collecting data at ~2 TeV center of mass energy. This increase in the energy was critical to produce pairs of the heaviest elementary particle known and last quark of the Standard Model, the top quark. As protons are made of quarks and gluons not only high center of mass energy is critical, but large number of interactions or the luminosity of the collider. In order to discover top quark luminosities of ~$10^{31}$ cm$^{-2}$ sec$^{-1}$ have been achieved by producing record large number of antiprotons in a specially designed antiproton source.

Tevatron high energy and luminosity as well as two modern detectors provided an opportunity to discover the top quark. Reconstructed mass for events used by the CDF and D0 experiments to discover the top quark is presented on Fig. 4. [6]. Tevatron led the energy frontier for twenty five years producing over 1000 publications covering all areas of particle physics from searches for new particles and interactions to high precision measurements of the Standard Model parameters.

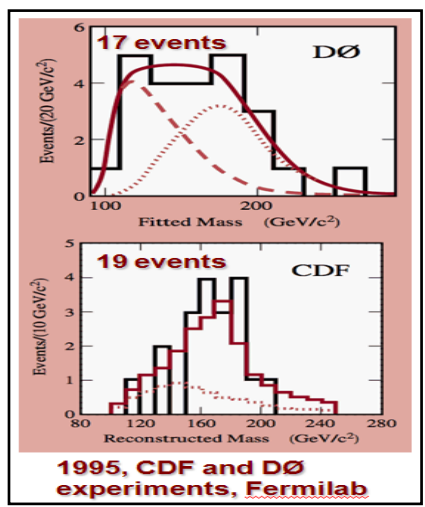

**Fig. 4.** Discovery of the top quark by CDF and D0 collaborations at the Tevatron.

## 2.4 To Even Higher Energies – Large Hadron Collider

In order to get to even higher colliding energies one of the main challenges became luminosity of the colliders. Partonic cross sections are decreasing inversely to the square of the energy and production of very large number of antiprotons and their cooling became a limiting factor. As a result collisions of protons with protons became better option for multi-TeV colliders while in comparison to ISR new machines combined acceleration and storage of the beams in the same ring.

Two hadron collider projects started in 1980's: UNK project with 6 TeV collision energy (near Serpukhov) and SSC project with 40 TeV collision energy (near Dallas). Both projects in part due to their substantial costs were not completed.

In 1994 LHC project at CERN received approval to build 14 TeV proton-proton collider with luminosity in the range of $10^{33}$–$10^{34}$ cm$^{-2}$ sec$^{-1}$. This collider was built over 15 years and since 2009 became the energy frontier collider.

High energy and luminosity of the LHC already provided fundamental discovery of the Higgs boson – last particle predicted by the Standard Model [3]. Excellent performance of both accelerator and general purpose detectors over last three years was the key reason for the success. Figs. 5a and 5b present plots of the Higgs boson discovery in the effective mass spectra of two photons. This discovery is a confirmation that progress in high energy physics is always related to developments in accelerators, detectors and analysis methods.

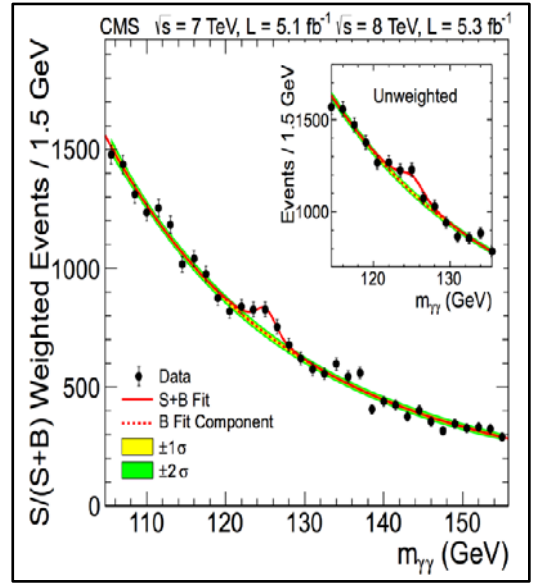

**Fig. 5b.** Discovery of the Higgs boson at the LHC – CMS collaboration.

## 3 Versatility of Hadron Colliders

While main goal of the energy frontier hadron colliders is the search for new heavy particles and interactions, very high luminosity and various quantum numbers of colliding partons provide an opportunity for extremely wide range of measurements. One of very productive areas of studies at the Tevatron collider is related to the particles containing b-quarks. For example, large number of new b-baryons, including Cascade, Sigma and Omega, have been discovered at the Tevatron as presented on Fig. 6.

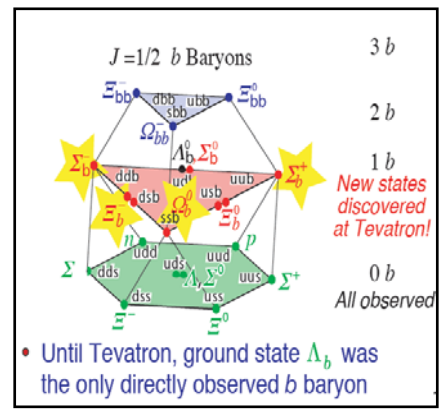

**Fig. 6.** Discovery of heavy baryons at the Tevatron.

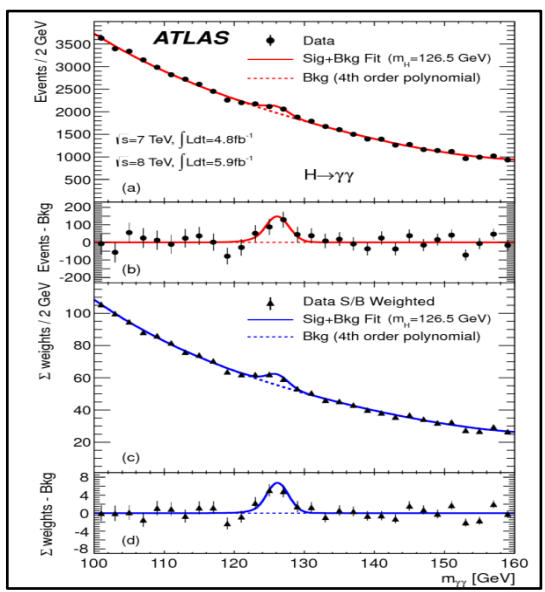

**Fig. 5a.** Discovery of the Higgs boson at the LHC – ATLAS collaboration.

Large samples of produced particles, including millions of top quarks at the LHC, provide an opportunity to measure fundamental properties of these particles with unprecedented precision. Top quark mass, now known to ~0.5% precision, is the most accurate quark mass measurement among all known quarks.

# 4 Detectors and Computing for Hadron Colliders

In order to address important physics topics and to fully exploit collisions provided by high energy and luminosity hadron colliders particle detectors have seen fantastic developments over past forty years. Without these developments none of the physics results observed would be possible.

One of the most serious challenges at a hadron collider is selection of "small" number of interesting events to be written to tapes out of 10's of MHz rate of interactions in the center of the detectors. In order to select most interesting events, for example with potential Higgs boson production and decay, multi-level trigger systems have been developed. At the first level trigger events are very quickly analyzed based on a generic information about topology and only those satisfying spesific criteria are sent to the next trigger level for further analysis. Up to three trigger levels are used in modern experiments providing efficient and dead-time free selection of 100's of events per second which are written to tapes for off-line analysis.

Adoption of silicon detectors with ~0.1mm resolution for tracking of charged particles (including triggering) revolutionized collider experiments providing an opportunity to tag decays of c- and b-quarks which propagate a few millimetres before decay. Selection of jets from b-quarks based on their decay products was the main reason for observing evidence for Higgs boson production with decay to b-quarks at the Tevatron.

Since experiments at the ISR it was realized that collider experiments have to cover close to 4π geometry, to have excellent energy and position resolution, been radiation hard to survive large fluxes of particles produced in the interaction region as well as to be well shielded from the accelerator created background particles. These requirements stimulated development of detectors with 100's of millions of channels, large sizes and weight. Cross section of the CMS detector at the LHC presented on Fig. 7 demonstrates complexity and versatility of the modern collider experiment.

Amount of data from 100's of millions of channels hadron collider detectors is enormous. Tevatron experiments data set is around 10 Petabytes and LHC experiments data sets are even larger. Analysis of such large data sets and verification of obtained results requires both modern computing systems, including GRID, as well as advanced particle detection algorithms. Hadron colliders experiments are leading in both of these activities providing science community with new ideas and resources.

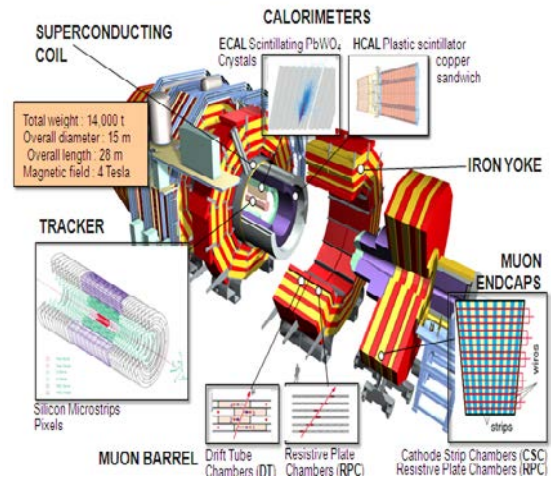

**Fig. 7.** CMS detector at the LHC.

# 5 Future Hadron colliders

In medium term LHC collider will lead the energy frontier. It is expected that LHC energy will reach ~14 TeV in 2014 and will provide ~300 $fb^{-1}$ of data by early 2020's. Increase in energy to ~14 TeV will provide an excellent opportunity to search for even heavier particles, for example, reaching ~5 TeV masses for hypothetical Z' and W' bosons. It will also provide large samples of top quarks and Higgs bosons to measure their properties with extremely high precision.

In ~2022 upgrades for an order of magnitude higher luminosity for both LHC accelerator and detectors are planned. Such upgrade is expected to provide ~3000 $fb^{-1}$ data set. While data taking conditions (such as number of interactions per crossing) will be rather challenging, high luminosity will provide an opportunity to study many rare processes and extend search for physics beyond Standard Model.

Beyond LHC proton-proton colliders remain the only practical option to continue increase in the collision energy which is critical to study even smaller size objects, including intrinsic structure of currently known elementary particles, and to create and study objects with masses in the multi-TeV range. For example, 100 TeV proton-proton collider will provide an opportunity to study distances up to $10^{-19}$ cm an order of magnitude better than achievable today.

There are two main options under study to increase energy of a hadron collider. One is development of even higher field bending magnets. Magnets with $Nb_3Sn$ superconducting coils can reach fields of ~15 T which is almost a factor of two above LHC magnets. Another area of development is construction of a long circular tunnel with length of ~100 km. With modern tunnelling technology such construction is becoming feasible, while reliability and servicing of such large accelerator will have to be developed.

In Table 2 parameters of Very Large Hadron Collider (VLHC) developed at Fermilab [7] are presented. It could reach energies close to 200 TeV using existing and proven technologies.

**Table 2.** Parameters of proton-proton VLHC collider.

| | Stage 1 | Stage 2 |
|---|---|---|
| Total Circumference (km) | 233 | 233 |
| Center-of-Mass Energy (TeV) | 40 | 175 |
| Number of interaction regions | 2 | 2 |
| Peak luminosity (cm$^{-2}$s$^{-1}$) | $1 \times 10^{34}$ | $2.0 \times 10^{34}$ |
| Luminosity lifetime (hrs) | 24 | 8 |
| Injection energy (TeV) | 0.9 | 10.0 |
| Dipole field at collision energy (T) | 2 | 9.8 |

## 6 Hadron Collider Physics Symposium

Hadron Collider Physics symposium (HCP) first conference was in 1979 in Paris and called "Topical Workshop on Forward Production of High-mass Flavors at Collider Energies". Twenty three HCP conferences proceeded since that time. The topics and format of the conferences was always adjusting to cover most interesting topics of hadron collider physics. During periods of intense activities, like during peak years of SPS operation, Tevatron 1990's and 2000's runs and early LHC years the conferences were called every year with about once in two years in between. Conferences covered wide range of countries, including Europe, USA, India and Japan indicative of diverse community involved in the hadron collider physics. These conferences provided extremely productive discussions and interactions between experimentalists and theorists involved in hadron colliders physics with many exciting results presented for the first time.

While World Wide Web was invented at CERN well after first of the HCP conferences, proceedings and talks from the conferences are available on the Web. Many of them could be found on the site of the last HCP symposium held in Kyoto (Japan) in November of 2012 [8].

## 7 Conclusions

For over forty years hadron colliders are leading the energy frontier providing major contributions to the particle physics including discovery of the W and Z bosons, top quark and Higgs boson. Standard Model has been completed with all expected particles discovered as indicated on Fig. 8.

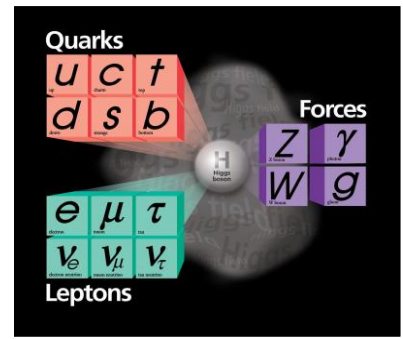

**Fig. 8.** Standard Model particles.

Progress in hadron colliders stimulated fantastic developments in particle detectors, trigger systems, particle detection algorithms and large data volume computing. In the short and medium terms analysis of data from the Tevatron and LHC and higher energy and luminosity LHC data set will provide experimental data to develop Standard Model and search for physics beyond the Standard Model. In a longer term proton-proton colliders remain the only practical way to extend energy frontier reaching ~30-100 TeV center of mass energy to study distances to ~10$^{-19}$ cm. Larger diameter accelerator rings and/or higher field magnets are the key aspects for success.

Since 1979 Hadron Collider Physics symposium provided excellent venue for experimental and theoretical physicists to present and discuss most interesting topics of the collider physics from the discoveries of W and Z bosons and top quark to Higgs boson discovery in depth discussed at the Hadron Collider Physics symposium in 2012.

## References

1. The CDF and D0 Collaborations, Phys. Rev. D **86**, 092003 (2012).
2. V. Shiltsev, Phys. Usp. **55** (2012) 965-976.
3. G. Aad *et al.* (ATLAS Collaboration), Phys. Lett. B **716**, 1 (2012); S. Chatrchyan *et al.* (CMS Collaboration), Phys. Lett. B **716**, 30 (2012).
4. The CDF and D0 Collaborations, arxiv:1204.0042v2.
5. UA1 Collaboration, Phys. Lett. B **122** (1983) 103-116; UA1 Collaboration, Phys. Lett. B **126** (1983) 398-410.
6. The CDF Collaboration, Phys. Rev. Lett. **74**, 2626 (1995); the D0 Collaboration, Phys. Rev. Lett. **74**, 2632 (1995).
7. VLHC Design Study Group, Design study for a staged very large hadron collider, FERMILAB-TM-2149.
8. XXIII HCP Conference, Kyoto (Japan), http://www.icepp.s.u-tokyo.ac.jp/hcp2012/M.